\documentclass[10pt]{article}

\usepackage{amsmath}
\usepackage{amssymb}

\usepackage{graphicx}

\usepackage{cite}

\usepackage{color} 

\usepackage[labelfont=bf,labelsep=period,justification=raggedright]{caption}

\makeatletter
\renewcommand{\@biblabel}[1]{\quad#1.}
\makeatother

\date{}

\begin{document}

\begin{flushleft}
{\Large
\textbf{Biological Averaging in RNA-Seq}
}
\\
Surojit Biswas$^{1,\ast}$, 
Yash N. Agrawal$^{2}$, 
Tatiana S. Mucyn$^{2}$,
Jeffery L. Dangl$^{2,3,4,5,6}$,
Corbin D. Jones$^{2,3}$
\\
\bf{1} Department of Statistics and Operations Research, University of North Carolina, Chapel Hill, North Carolina, United States of America
\\
\bf{2} Department of Biology, University of North Carolina, Chapel Hill, North Carolina, United States of America
\\
\bf{3} Carolina Center for Genome Sciences, University of North Carolina, Chapel Hill, North Carolina, United States of America
\\
\bf{4} Howard Hughes Medical Institute, University of North Carolina, Chapel Hill, North Carolina, United States of America 
\\
\bf{5} Curriculum in Genetics and Molecular Biology, University of North Carolina, Chapel Hill, North Carolina, United States of America
\\
\bf{6} Department of Microbiology and Immunology, University of North Carolina, Chapel Hill, North Carolina, United States of America
\\
$\ast$ E-mail: sbiswas@live.unc.edu
\end{flushleft}

\section*{Abstract}
RNA-seq has become a \emph{de facto} standard for measuring gene expression. Traditionally, RNA-seq experiments are mathematically averaged -- they sequence the mRNA of individuals from different treatment groups, hoping to correlate phenotype with differences in arithmetic read count averages at shared loci of interest. Alternatively, the tissue from the same (or more) individuals may be pooled prior to sequencing in what we refer to as a biologically averaged design. As mathematical averaging sequences all individuals it controls for both biological and technical variation; however, is the statistical resolution gained always worth the additional cost? To compare biological and mathematical averaging, we examined theoretical and empirical estimates of statistical efficiency and relative cost efficiency. Though less efficient at a fixed sample size, we found that biological averaging can be more cost efficient than mathematical averaging, especially if biological variation is large and biologically averaged individuals can be pooled evenly. With this motivation, we developed a differential expression classifier, ICRBC, that can detect alternatively expressed genes between biologically averaged samples. In simulation studies, we found that biological averaging and subsequent analysis with our classifier performed comparably to existing methods, such as ASC, edgeR, and DESeq, especially when individuals were pooled evenly and less than 20\% of the regulome was expected to be differentially regulated. In two technically distinct mouse datasets and one plant dataset, we found that our method was over 87\% concordant with edgeR for the 100 most significant features. While biological averaging cannot provide the same statistical resolution as a well replicated mathematically averaged experiment, it may sufficiently control biological variation to a level that differences in gene expression may be detectable. In such situations, ICRBC can enable reliable exploratory analysis at a fraction of the cost, especially when interest lies in the most differentially expressed loci.


\section*{Introduction}
RNA-Seq \cite{Mortazavi2008a} is a popular alternative to microarray based methods for measuring gene expression \cite{Shendure2008}. RNA-seq quantitatively measures gene expression as counts, and typically involves isolating a subject's mRNA, converting to cDNA, and sequencing. Sequencing reads are then computationally ``mapped'' to loci of interest (e.g. genes or exons), and the number of reads associated with each loci is stored in a $p$-loci by $n$-individuals matrix \cite{Wang2009, Mortazavi2008a}. Matrices built from individuals representing phenotypically different populations may then be compared in order to correlate differences in gene expression with phenotype. 

Most current methods for detecting differential expression utilize the mathematical average of a gene's read counts across sequenced samples from the same population as a point estimate for its expression \cite{Robinson2007, Anders2010, Li2011, Wang2010, Wu2013}. We refer to these traditional designs, in which each individual's mRNA is sequenced, as \emph{mathematically averaged} experiments. Because multiple samples are sequenced, mathematical averaging allows the researcher to control for not only biological variation, but also technical variation that arises from the measurement process \cite{Auer2010}.  While technical variation is typically too large to ignore \cite{McIntyre2011}, biological variation is typically more pronounced \cite{Bullard2010}.

In contrast, one may also consider designs in which tissue from each individual is pooled prior to mRNA isolation and sequencing.  Since the pool represents an average of biological samples, we refer to these designs as \emph{biologically averaged} experiments. Biological averaging has received considerable attention in the microarray community \cite{Kendziorski2005, Zhang2005, Zhang2007}, where it's believed most effective when 1) biological variation exceeds technical variation and 2) many individuals can be pooled. In fact, Zhang (2007) \cite{Zhang2007} show even when biological and technical variation are equal in magnitude, arraying ten pools of ten individuals each affords 90\% of the statistical power at 10\% of the cost of a study that arrays every individual. Surprisingly, there has been little statistical treatment of biological averaging the RNA-Seq literature. Due to sequencing costs and experimental scale, several studies applied RNA-Seq to pooled \cite{Zenoni2010, Mills2013, Greenwald2012} or unreplicated \cite{Cheng2010, Marti2010, Cui2010} samples, but employed statistical tests (e.g. Fisher's exact test \cite{Auer2010}) and methods (e.g. DEGSeq \cite{Wang2010}) that only adequately model technical variability. 

It's clear that both mathematical and biological averaging control for subject-to-subject variability and makes population level differences easier to detect \cite{Auer2010}; however, what are the statistical and monetary trade-offs associated with each design? By what metric can we suggest one design is ``better'' than the other? How do the number of individuals enrolled and pooling evenness affect the quality of biologically averaged gene expression estimates? To our knowledge, the empirical bayes routine, ASC, is the only method that is capabable of analyzing biologically averaged experiments\cite{Wu2010}; however, the authors of this method do not provide a statistical justification of biological averaging.

In this work, we consider the relative efficiency \cite{Hoel1971} of biological and mathematical averaging in estimating intrapopulation level expression levels. Additionally, we examine a modified objective -- the relative cost efficiency -- which also considers experimental cost. We show on three biologically and technically varied datasets \cite{Katz2010, Cumbie2011, Bottomly2011} that biological averaging can be statistically more cost efficient than mathematical averaging, and that it may be particularly useful for exploratory analysis if interpopulation similarity is high, multiple individuals are enrolled, and individuals are pooled relatively evenly.  For such scenarios, we present an iterative confidence-region based classifier (ICRBC) to detect differentially expressed genes in biologically averaged experiments. We provide results from two simulation experiments and real data to show our classifier provides reasonable results, especially when interest lies in the most significant loci.

\section*{Statistical concerns}

For any significance test, a random variable's expected value and variance under an assumed null distribution are required. In this section, we consider statistical practicalities of estimating the expected value and variance of a gene's population level expression under mathematically and biologically averaged designs. We consider their relative cost efficiency -- a measure of statistical power gained per dollar -- and provide a motivating example from real data that suggests the potential utility of biological averaging.

\subsection*{Preliminaries}
\begin{figure}[!ht]
\begin{center}
\includegraphics[width=6in]{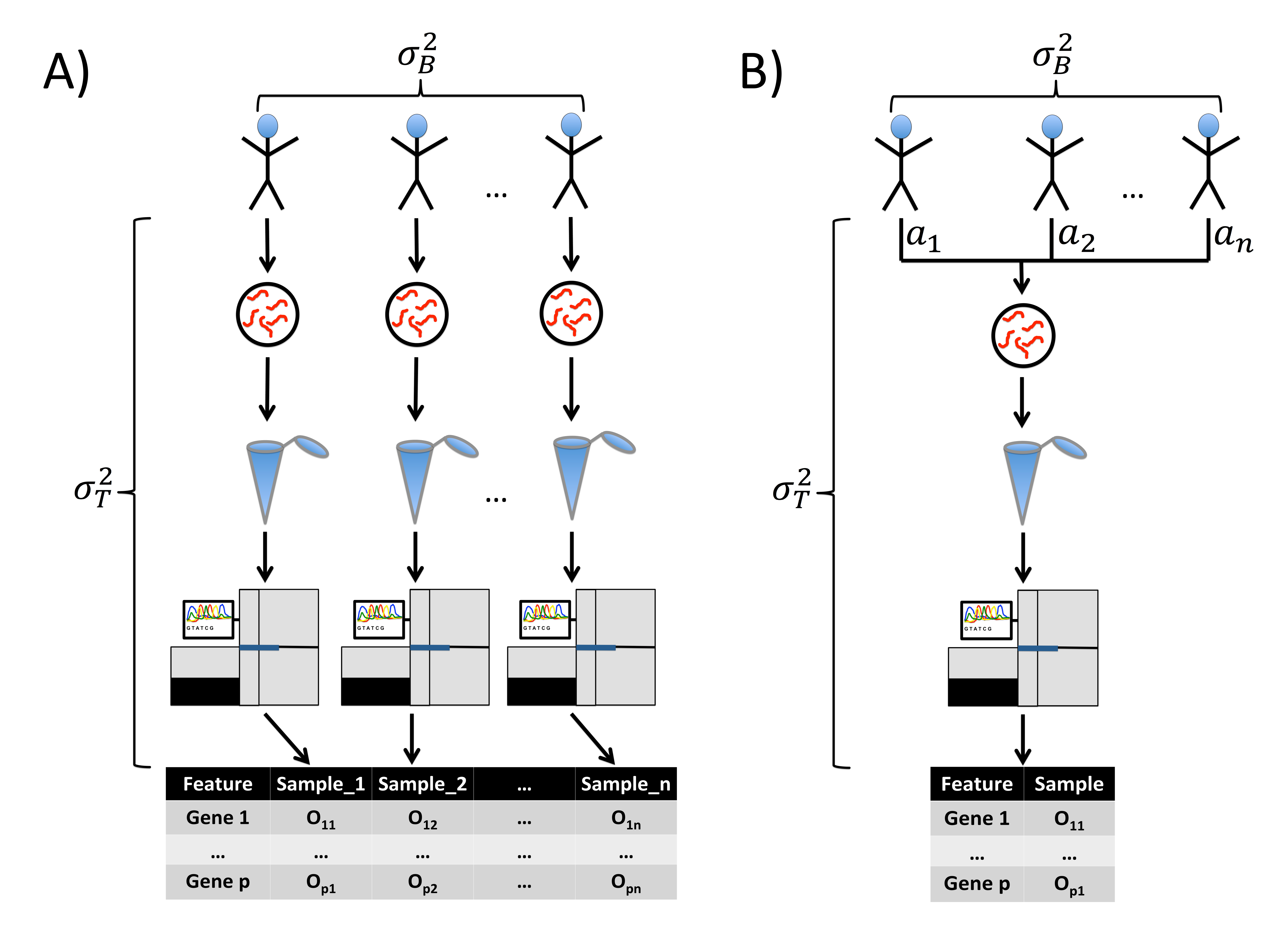}
\end{center}
\caption{
{\bf Schematic of a A) traditional mathematically averaged experiment, and B) a biologically averaged one.}  Biological variation between individuals at a particular locus is denoted by $\sigma^2_B(g)$, and the technical variation associated with the measurement process is given by $\sigma^2_T(g)$. For biologically averaged experiments, the coefficients $a_1, \ldots a_n$ denote the pooling proportions of the each individual.
}
\label{fig1}
\end{figure}

Figure \ref{fig1}A illustrates a typical workflow for a mathematically averaged design, in which mRNA is isolated, converted to cDNA, and sequenced for each of $n$ individuals \cite{Auer2010}. Thereafter, sequencing reads are mapped to loci of interest and the final output is a $p$-loci by $n$-individuals count matrix, $O$.  The entry, $O_{ij} \in \mathbb{N}_0$, of this count table, denotes the number of reads that map to loci $i$ in individual $j$. Figure \ref{fig1}B depicts a biologically averaged design in which tissue from each individual is instead pooled into a single sample prior to mRNA isolation and sequencing. The proportional representation of the $i^{th}$ individual in the pool is denoted by $a_i \in [0,1]$.  We assume that individuals are independent and identically distributed (IID) with respect to their expression profiles.

Assuming the measurement process (mRNA isolation, conversion to cDNA, sequencing, and the computational mapping of reads) is independent of all individuals, the total variance in read count at a particular locus is  the sum of biological and technical variance \cite{Anders2010}. Biological variation, denoted by $\sigma^2_B(g)$, is the natural variation in expression level for gene $g$ found among individuals in a given population. Technical variation, given by $\sigma^2_T(g)$, is the additional variation in the estimated expression level of gene $g$ introduced by the measurement process. 

\subsection*{Relative Cost Efficiency}
 
\subsubsection*{Theory}

The relative efficiency between two estimators or measurement procedures is given by the ratio of their variances \cite{Hoel1971}. If the variance of one procedure is larger than the other for a fixed sample size, $n$, then the procedure is less efficient and necessarily requires more data to detect differential expression at a prespecified level of confidence (i.e. it has a lower signal-to-noise ratio). \\

Let $d_m(g,O) = \frac{1}{n}\sum_{i=1}^{n}O_{gi}$ denote the point estimate of gene $g$'s expression from a mathematically averaged experiment, and let $d_b(g,O) = O_g$ denote the biologically averaged estimate. The relative efficiency of these estimators is given by (see derivation in Section 1 of SI): 
\begin{align*}
\textrm{eff}_{rel}(d_b, d_m)  & = \frac{\textrm{V}[d_b(g, O)]}{\textrm{V}[d_m(g, O)]} \\
& = \frac{ \sigma_T^2(g) + \sigma_B^2(g) \epsilon }{ \frac{1}{n} (\sigma_T^2(g) + \sigma_B^2(g)) } > 1.
\end{align*}

Here $\epsilon$ is a measure of pooling evenness and is bounded between $1/n$ and $1$. 

Notice that $\textrm{eff}_{rel}$ is always greater than $1$, implying that a biologically averaged experiment is less efficient for fixed $n$. This is expected because a biologically averaged design cannot control for technical variation through replication of the sequencing process. However, it may be useful to also consider the \emph{relative cost efficiency}, $\textrm{eff}_{rc}$, which proportionally penalizes the relative efficiency by experimental cost. 

If we let $q$ denote the cost of obtaining a tissue sample from an individual, and $s$ the cost of preparing and sequencing a single library, then the relative cost efficiency is given by (see derivation in Section 2 of SI),
\begin{align*}
\textrm{eff}_{rc}(n | d_b, d_m) & = \frac{\textrm{cost}(d_b)}{\textrm{cost}(d_m)} \times \frac{\textrm{V}[d_b(g, O)]}{\textrm{V}[d_m(g, O)]} \\
& =  \frac{nq + s}{q+ s} \times \frac{ \sigma_T^2(g) + \sigma_B^2(g) \epsilon }{ \sigma_T^2(g) + \sigma_B^2(g) }
\end{align*}
Intuitively, this objective favors the method that best manages the trade off between experimental cost and estimation precision. Like before, a value for $\textrm{eff}_{rc} > 1$ suggests mathematical averaging is a more desirable design; however, because experimental cost is also considered, this objective quite literally quantifies ``bang for the buck.'' 

Note that previous derivations for $\textrm{eff}_{rc}$ are invariant with respect to any distributional assumptions about read counts. If we now make the standard assumption that read counts follow a $\textrm{Negative-Binomial}(\mu, \alpha)$  distribution \cite{Robinson2007, Anders2010}, then under the NB2 parameterization, $\sigma^2_T(g) = \mu$ and $\sigma^2_B(g) = \alpha \mu^2$, where $\mu$ and $\alpha$ denote gene $g$'s expected expression level and the dispersion parameter, respectively \cite{Anders2010}. Thus, 
\begin{align*}
\textrm{eff}_{rc}(n, \mu | d_b, d_m) & =  \frac{nq + s}{q+ s} \times \frac{ \mu+ \alpha \mu^2 \epsilon }{ \mu + \alpha \mu}
\end{align*}

Finally, instead of conditioning only on the expression level of a single gene, we may average over all expression levels by considering the expected relative cost efficiency (ERCE),

\begin{align*}
\textrm{E}[\textrm{eff}_{rc}(n, \mu | d_b, d_m) ]& =  \left(\frac{nq + s}{q + s} \right) \times \\
&\int_{\mu \in \mathbb{R}_+} \frac{ \mu + \alpha \mu^2 \epsilon }{ \mu + \alpha \mu^2 } f(\mu) \hspace{1mm}\textrm{d}\mu
\end{align*}

were $f(\cdot)$ is a probability density function. 

\subsubsection*{Pooling evenness}
In the efficiency expressions above, $\epsilon = \sum_{i=1}^n a_i^2$, and is a measure of how evenly samples are pooled in a biologically averaged design. It is uniquely minimized when $a_i = 1/n \; \forall_{i}$ (a perfectly even pool), and uniquely maximized when $a_i = 1$ and $a_j = 0 \; \forall_{j \ne i}$ (a perfectly uneven pool -- only one individual is effectively included). 

When among $n$ individuals only a single individual's tissue is included in the pool ($\epsilon = 1$), the integral in the expression for the ERCE evaluates to unity. In this case, the expected relative cost efficiency grows linearly in $n$, and remains greater than 1, suggesting that mathematical averaging is the better design. This is intuitively reasonable because a stable (low-variance) estimate of a gene's expression is difficult to obtain with a single individual. 

By contrast, consider when all $n$ individuals are equally represented in the pool ($\epsilon = 1/n$). In this case, the relationship between the expected relative cost efficiency and $n$ is non-trivial. It is easily shown that when $\epsilon = 1/n$,  $\textrm{eff}_{rc}(n, \mu | d_b, d_m)$ is concave for all $n > 1$, and uniquely minimized by $n^* = (\sigma_B(g)/\sigma_T(g))\sqrt{s/q}$; however whether the ERCE is less than 1 (biological averaging is more cost effective) will depend on the relative magnitudes of not only tissue isolation and sequencing costs, but also technical and biological variation. Additionally, as we will see, pooling evenness plays an influential role.

\subsubsection*{Empirical motivation}
Because experimental costs, pooling evenness, and the magnitudes of technical and biological variation depend on  experimental design and organism, definite trends in the ERCE (i.e. whether it is above or below 1) are not readily abstractable. Moreover, the ERCE is ultimately a function of $n$, the number of individuals to be included in an RNA-seq experiment. Given that $\textrm{eff}_{rel}$ is concave in $n$, the ERCE will also be concave in $n$, but does it ever go below 1? If so, for what $n$? Most importantly, what are the practical implications of its behavior?

To answer these questions we looked at three RNA-Seq datasets from two technically distinct mouse experiments \cite{Katz2010, Bottomly2011} and one plant experiment \cite{Cumbie2011}. Because these datasets are used frequently hereafter, we introduce each below and mention their relevance to this work. We refer to each dataset by the last name of the first author. \\

\noindent\textbf{Dataset description}\\
\textbf{Cumbie} -- Cumbie \emph{et al.} developed the fully integrated computational pipeline GENECounter for the analysis of RNA-Seq data \cite{Cumbie2011}. To test their package, they inoculated the leaves of \emph{Arabidopsis thaliana} plants with either MgCl$_2$ (control) or an avirulent $\Delta$\emph{hrcC} mutant strain in order to assess whether their analysis pipeline could detect differentially expressed genes known to be involved in the well studied plant Hypersensitive Response (HR) to avirulent pathogens \cite{Jones2006}. Their dataset consists of three wildtype and three mutant biological replicates, and provides an organismal contrast to the two mouse datasets detailed below. 

\textbf{Katz} -- Katz \emph{et al.} developed MISO, a statistical model designed to detect differential isoform expression \cite{Katz2010}. In addition to investigating the possible roles of the splicing factor hnRNP H in isoform regulation and alternative polyadenylation, they examined the effects of read pairing and library insert length on MISO's ability to identify differentially regulated isoforms between normal mouse myoblasts and myoblasts depleted of the splicing factor CUGBP1. Their data consists of two biological replicates of normal myoblasts and two biological replicates of  CUGBP1 depleted myoblasts. Because each biological replicate within each population was prepared with a different insert length, this dataset is technically more varied than the Cumbie and Bottomly datasets.

\textbf{Bottomly} -- Bottomly \emph{et al.} compared the concordance between RNA-Seq and two microarray platforms in detecting differential striatal gene expression between C57BL/6J (B6) and DBA/2J (D2), two commonly used inbred mouse strains in neuroscience research \cite{Bottomly2011}. Their dataset consists of 10 B6 and 11 D2 striatal tissue samples. Their large sample size enables us to compare the marginal benefit of sequencing many samples to sequencing only a single, biologically averaged sample. 
\\

\noindent\textbf{Expected relative cost efficiency curves}\\
Using edgeR, we obtained a boostrapped estimate, $\hat{\alpha}$, of each dataset's dispersion parameter. Using each study's method description, we additionaly estimated the cost of tissue isolation, $q$, and the cost of sample preparation and sequencing to each study's median depth, $s$. Table \ref{table1} summarizes these estimates (see Section 2.2 in the SI for a derivation of our estimates). Using the median count profile of all control (e.g. wildtype) samples as an exemplar for the ``typical'' individual, we examined how nonparametric estimates of the ERCE would behave for each study if more hypothetical individuals, similar to the exemplar, were enrolled. Section 2.1 in the Supplementary Information provides a complete description of this nonparametric fitting procedure.

\begin{figure}[!ht]
\begin{center}
\includegraphics[width=6in]{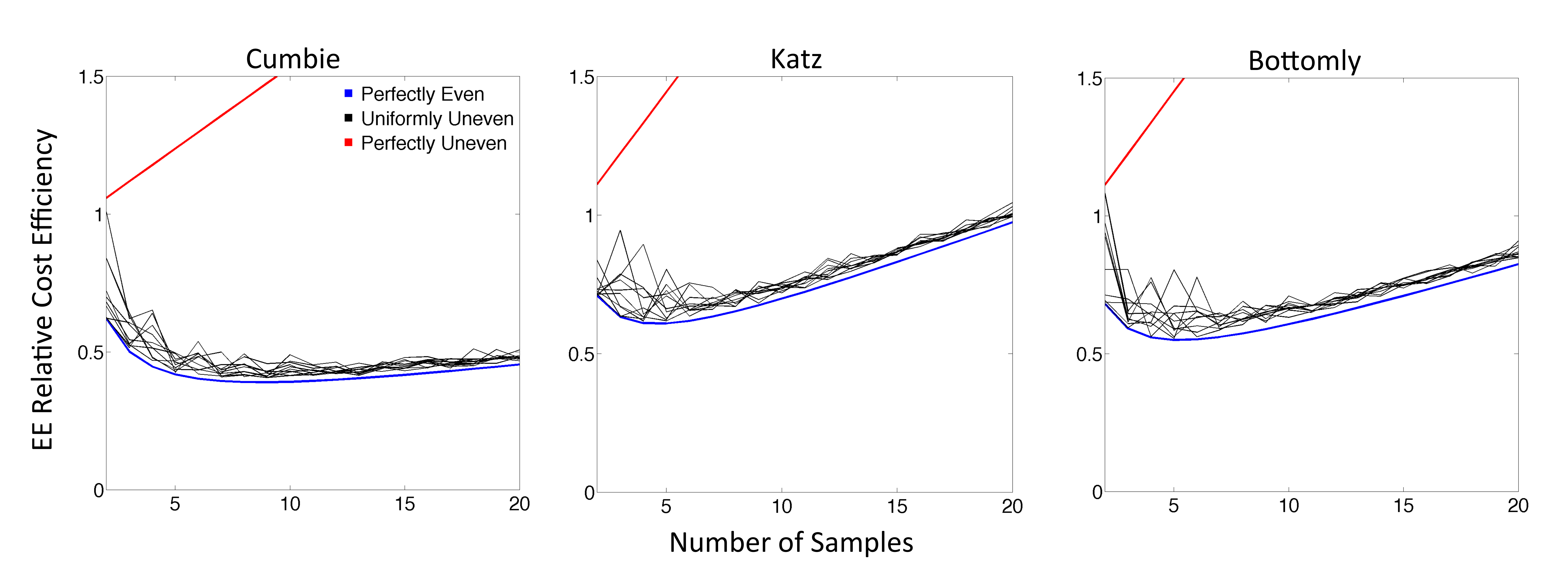}
\end{center}
\caption{
{\bf Empirical estimates of the ERCE function as a number of individuals for the Cumbie, Katz, and Bottomly datasets. }  Mathematically averaged designs are naturally more sensible than perfectly unevenly pooled biological ones; however, even for uniformly uneven pooling, the estimated ERCE remains below 1 for the first twenty individuals, suggesting that biological averaging is the more cost efficient design in these datasets.
}
\label{fig2}
\end{figure}

Figure \ref{fig2} illustrates our empirical estimates of the ERCE as a function of the number of hypothetical individuals enrolled. Because the ERCE is sensitive to how evenly individuals are pooled during biological averaging, we considered three pooling types: 1) perfectly even ($\epsilon = 1/n$), 2) perfectly uneven ($\epsilon = 1$), and 3) randomly generated uniformly uneven pools, designed to represent human error in pooling a biologically averaged library (see Figure 2 in SI). 

As expected, when pooling is completely uneven, mathematical averaging handily outperforms biological averaging. On the other hand, the ERCE for a perfectly even pool remains below 1 for $2 \leq n \leq 20$ for all three datasets, and illustrates the idealized case of zero pooling error in a biologically averaged design.

Surprisingly, the ERCE trend is much the same for uniformly uneven pools. There is considerable variability in the ERCE when only a few individuals are pooled under this scheme; however, this variance stabilizes quickly and shortly after the minimum ERCE, which seems to occur typically between 3 and 10 individuals.

Interestingly, the uptrend in the ERCE for Cumbie is attenuated relative to the uptrends for Katz and Bottomly. The slower rise for Cumbie is partially attributable to the increased cost of sequencing individuals to a depth of 5 million reads. However, further sensitivity analyses via permutation test of the dispersion and cost estimate effects on the ERCE revealed that the most influential factor was the higher biological dispersion, which is roughly two-fold greater in the Cumbie data set.  In theory, this is likely attributable to the difference in organism (plant versus mouse).

Taken together, these results suggest that biological averaging can provide a 1.3-2.2 fold signal-to-noise increase per dollar, thus motivating its potential use as an experimental design especially in experimental systems with large biological variation.

\section*{Methods}

While it may be a more cost efficient solution in some cases, biological averaging presents unique challenges for differential expression expression analysis. Biological averaging may be combined with mathematical averaging by sequencing multiple pools of individuals. However, in this work we assume individuals from the same population are averaged into a single pool. Because there is only sequenced replicate per condition, independent variance estimates at each locus cannot be readily obtained. However, if biological variance is larger than technical variance \cite{Bullard2010} and a minority of genes are expected to be induced by a population condition \cite{Yeung2002}, variance estimates can be obtained by pooling information across loci \cite{Anders2010, Wu2010, Robinson2007}.   We now an iterative confidence region based classifier (ICRBC) that is capable of differential expression analysis in biologically averaged experiments. 

\subsection*{Modeling}

The Negative-Binomial distribution has been commonly applied to RNA-seq data as an overdispersed Poisson model \cite{Robinson2007, Anders2010, Cumbie2011}. Most frequently the NB2 parameterization is employed, in which $\textrm{E}[O_g] = \mu$ and $\textrm{V}[O_g] = \mu + \alpha \mu^2$ \cite{Hilbe2011}. The parameter, $\alpha$, is known as the dispersion parameter, and it signifies the amount of intrapopulation biological variation. Setting it equal to 0 recovers the original Poisson model \cite{Hilbe2011}. 

In this work, we assume intrapopulation read counts follow a NB2 parameterized, $\textrm{Negative-Binomial}(\mu, \alpha)$ distribution. Let $O_{:,1}$ and $O_{:,2}$ denote the observed read count vectors of two sequenced samples representing two different populations in a biologically averaged experiment. Let $M = \log_2(O_{:,2}) - \log_2(O_{:,1})$ and $A = (\log_2(O_{:,2}) + \log_2(O_{:,1}))/2$. This transformation is known as the MA-trasformation \cite{Smyth2003}.

\begin{figure}[!ht]
\begin{center}
\includegraphics[width=4in]{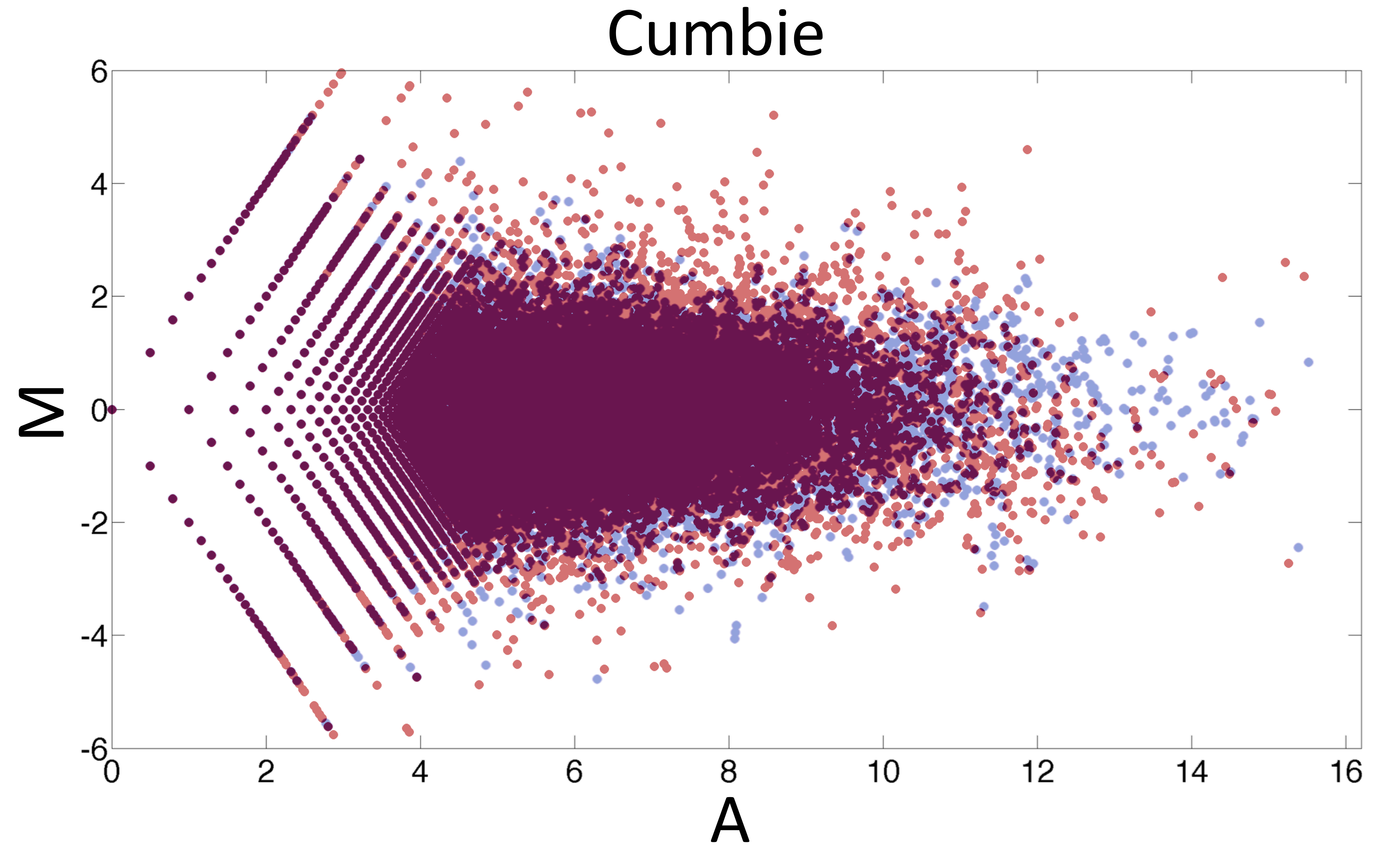}
\end{center}
\caption{
{\bf Overlayed MA plots of control ($\textrm{MgCl}_2$) vs. control (blue) and control vs. treatment ($\textrm{DC3000}_{\Delta \emph{hrcC}}$, red) samples from the Cumbie dataset. }  While the mean of $M$ is constant across $A$, considerable heteroscedasticity can be seen. In addition, near the superior edge the cloud of red points extends beyond the cloud of blue points, suggesting the loci these red points represent may be differentially expressed.
}
\label{fig3}
\end{figure}

Using delta method \cite{Casella2002}, it can be shown that $\log_2(O_{i,j})$ is approximately normally distributed given gene $i$ is not differentially expressed. Using this approximation, we show in Section 3 of the Supplemental Information that $M_{| A = a} \sim \textrm{Normal}(\theta, g(a))$, where $g(\cdot)$ is a continuous and smooth function of $a$. Note that while $\theta$ is invariant with respect to $a$, the variance of $M$ with respect to $A$ is not. The level mean and overall heteroscedasticity as functions of $A$ can be visually seen in Figure \ref{fig3}, where two MA plots of two different sample pairings from the Cumbie dataset have been overlaid.

\subsection*{Iterative Confidence Region Based Classification (ICRBC)}

Figure \ref{fig3} shows overlayed MA plots made from a control/control sample pairing (blue data points), where all genes are expected to follow a null distribution, and a control/$\Delta$\emph{hrcC} sample pairing, (red data points), where some genes may be differentially expressed. Near the superior edge of the point cloud (more positive $M$), there are red data points that extend vertically beyond the mass of blue data points, implying these red points may represent differentially expressed genes. In other words, these genes appear as ``outliers'' when compared to the larger mass of regularly expressed features.

\begin{enumerate}
\item[] \textbf{Algorithm 1:} ICRBC
\item \textbf{Input:} $O_{:,1}, O_{:,2}, \kappa$
\item \texttt{[}$a$\texttt{,} $m$\texttt{]} $=$ \texttt{maTransform(}$O_{:,1}$, $O_{:,2}$\texttt{)}
\item $ S = \{i : |m_i - \texttt{median(}m\texttt{)}| < 0.05\times \texttt{std(}m\texttt{)}\}$
\begin{enumerate}
\item[] \textbf{while} $S$ changing
\item $\hat{\theta} = \texttt{mean}(m(S))$
\item $\hat{g}(x) = \texttt{loess}\left(a(S), \left[m(S) - \hat{\theta} \right]^2\right)$
\item $z = \left(m - \hat{\theta} \right)/\sqrt{\hat{g}(a)} $ \;
\item $S = \{i : |z_i| < t_{1 - \kappa/2, n - 2} \} $\; 
\end{enumerate}
\item \texttt{[}$p$\texttt{,} $q$\texttt{] = pqVals(}$m$, $a$, $\hat{g}(x)$, $\hat{\theta}$\texttt{)}\;
\item \textbf{Return:} $S, q, p$
\label{icrbc}
\end{enumerate}


Algorithm 1 details an Iterative Confidence Region Based Classifier (ICRBC), which detects differentially expressed genes (the ``outliers'') from MA-transformed count data. During initialization, two input count vectors from two biological averaged experiments are MA-transformed, and a set of indices, $S$, is chosen to be indices of those loci with $m$-coordinates that slightly deviate from the median of all $m$-coordinates. While the elements of $S$ continue to change, the algorithm iteratively alternates between estimating parameters of the null model -- $g(a)$ and $\theta$ -- and estimating the index set of loci participating in the null model, $S$. 

Specifically, an estimate of $\theta$ is obtained by the sample mean of $m$-coordinates belonging to loci currently estimated to be null. The variance function, $g(a)$ is estimated by performing local regression of the squared null $m$-coordinate residuals onto the null $a$-coordinates. This effectively pools information across loci with similar expression levels, and in so doing, provides an unbiased estimate of the conditional $m$-coordinate variance. The local regression is done using weighted linear least squares and a $2^{nd}$ degree polynomial model \cite{Cleveland2013}. Subsequently, all $m$-coordinates are $z$-transformed (standardized), and the indices of those standardized coordinates that lie within $(\kappa/2, 1 - \kappa/2)$-percentiles of the $t$-distribution are set as the indices of null features. The $t$-distribution is used here because the standardizing mean and variance are estimated quantities. Throughout this work we use $\kappa = 0.01$.

The procedure concludes with significance testing of each loci using the final estimates of $\theta$ and $g(a)$. Feature $p$-values are calculated with respect to tail probabilities of the $t$-distribution, and $q$-values are obtained using the method of \cite{Benjamini1995}.

At its core, the ICRBC algorithm is simply a series of sequential hypothesis tests that serve as filters that let pass the loci that do not appear too extreme given a current null set, and screen out the loci that do. Alternatively, if the null index set $S$ is considered as auxillary or missing-data, then Algorithm \ref{icrbc} can be seen as an approximate hard-EM algorithm \cite{Dempster1977}, where the E-step involves estimating $S$ given $\theta$ and $g(a)$, and the M-step involves estimating $\theta$ and $g(a)$ given $S$. The approximation arises from estimating $g(a)$ nonparametrically, as opposed to maximizing the expected log-likelihood function with respect to the conditional distribution of $S$.

\section*{Results}

\subsection*{Baseline Comparisons}

Throughout this section, we compare ICRBC to edgeR \cite{Robinson2007}, DESeq \cite{Anders2010}, and ASC \cite{Wu2010}. The edgeR and DESeq models are both based on the Negative Binomial distribution, require at least two biological replicates, and pool information across loci of similar expression values in order to estimate the dispersion parameter. In a recent survey of eleven differential expression detection algorithms, the authors of \cite{Soneson2013} found that edgeR and DESeq most often performed the best and noted that edgeR tended to be more liberal in calling differential expression, whereas DESeq was more conservative.

The Analysis of Sequence Counts (ASC) is an empirical Bayes method for detecting differential expression in biologically averaged experiments. Like ICRBC, ASC estimates the expected null variation in expression levels between two samples by conditioning on average expression. However, instead of estimating this relationship nonparametrically as ICRBC does, ASC assumes in its prior that $\log_{10}$RPM expression values across loci follow a shifted exponential distribution.

\subsection*{Simulation Studies}

\subsubsection*{Biological versus Mathematical Averaging}

In this experiment, we explore how ICRBC and the baseline methods perform in simulated mathematically and biologically averaged samples in order to assess how ICRBC performs as a function of 1) the number of individuals and 2) pooling evenness. Additionally, we directly compare how ICRBC's differential detection rates on biologically averaged samples compare to edgeR, which uses all available sequenced replicates. 

\textbf{Data generation} -- The three control (MgCl$_2$) count vectors from the Cumbie dataset were mathematically averaged, gene-by-gene, and set to be the `normal' population level expression vector, $\mu$. A total of $p = 33672$ loci had non-zero expression levels after averaging. To simulate `normal' tissue samples, each of 10 $p$-long $\textrm{Gamma}(\alpha, 1/\alpha)$ random vectors (parameterized by shape and scale, respectively) were first multiplied element-wise with $\mu$ thereby producing 10 \emph{transcript vectors} with mean $\mu$ and variance $\alpha\mu^2$ \cite{Hilbe2011}. Transcript vectors represent transcript abundances in a tissue sample, and not read counts obtained after sequencing -- variability is entirely attributable to biological variation ($\alpha \mu^2$). The dispersion parameter, $\alpha$ was set to be 0.03.

Ten `abnormal' tissue samples were simulated by first randomly selecting 12.4\%  (4160) of the $33672$ genes to be differentially expressed. Differential expression was simulated by altering $\mu$ to $\mu^*$, where selected loci (entries of $\mu$) were modified the formula $\mu^{*}_i = U_{[1.6, 5]}\text{sign}\{U_{[0,1]} - 0.3 \}\sqrt{\alpha \mu_i^2}$. Here, $U_{[a,b]}$ denotes a $\textrm{Uniform[a,b]}$ random number. In words, approximately 70\% of the genes chosen to be differentially expressed were upregulated (the remaining were downregulated), and the degree of expression difference was set to be 1.6 to 5 standard deviations away from the true expression value. Note that the 1.6 to 5 standard deviation interval suggests considerable overlap between the null distribution and the alternative, thus making the classification task non-trivial. Subsequently, 10 `abnormal' transcript vectors were generated by multiplying 10 $p$-long $\textrm{Gamma}(\alpha, 1/\alpha)$ by $\mu^*$. The dispersion parameter, $\alpha$, for `abnormal' individuals was also set to 0.03. 

From the transcript vectors, we then proceeded to simulate the sequencing process to produce \emph{read count vectors}. To simulate a count table for a mathematically averaged design of $n$ individuals ($n \in \{1,2, \ldots, 10\}$), each of the 20 transcript vectors (10 `normal' and 10 `abnormal') were first multinomially sampled $\lceil U_{[2\times 10^6, 4 \times 10^6]} \rceil$ times to produce 20 read count vectors. The process of multinomial sampling simulates the sequencing process by adding technical variance of magnitude $\mu$ \cite{Casella2002, Mortazavi2008a, Robinson2007}. Subsequently, for each population $n$ read count vectors were concatenated to produce a $33672 \times 2n$ count table.

To simulate a count table for a biologically averaged design of $n$ individuals, $n$ transcript vectors from each population were averaged with either even weights (normal arithmetic average) or uneven weights. This averaged transcript vector was then multinomially sampled $\lceil U_{[2\times 10^6, 4 \times 10^6]} \rceil$ times, to produce a single read count vector for each population. Uneven averaging was done such that a single individual always had an pooling weight greater than 0.4 (see Figure 3 in SI for exact pooling weights). This provides for a substantially more uneven pool than is likely to occur in practice.

\begin{figure}[!ht]
\begin{center}
\includegraphics[width=4in]{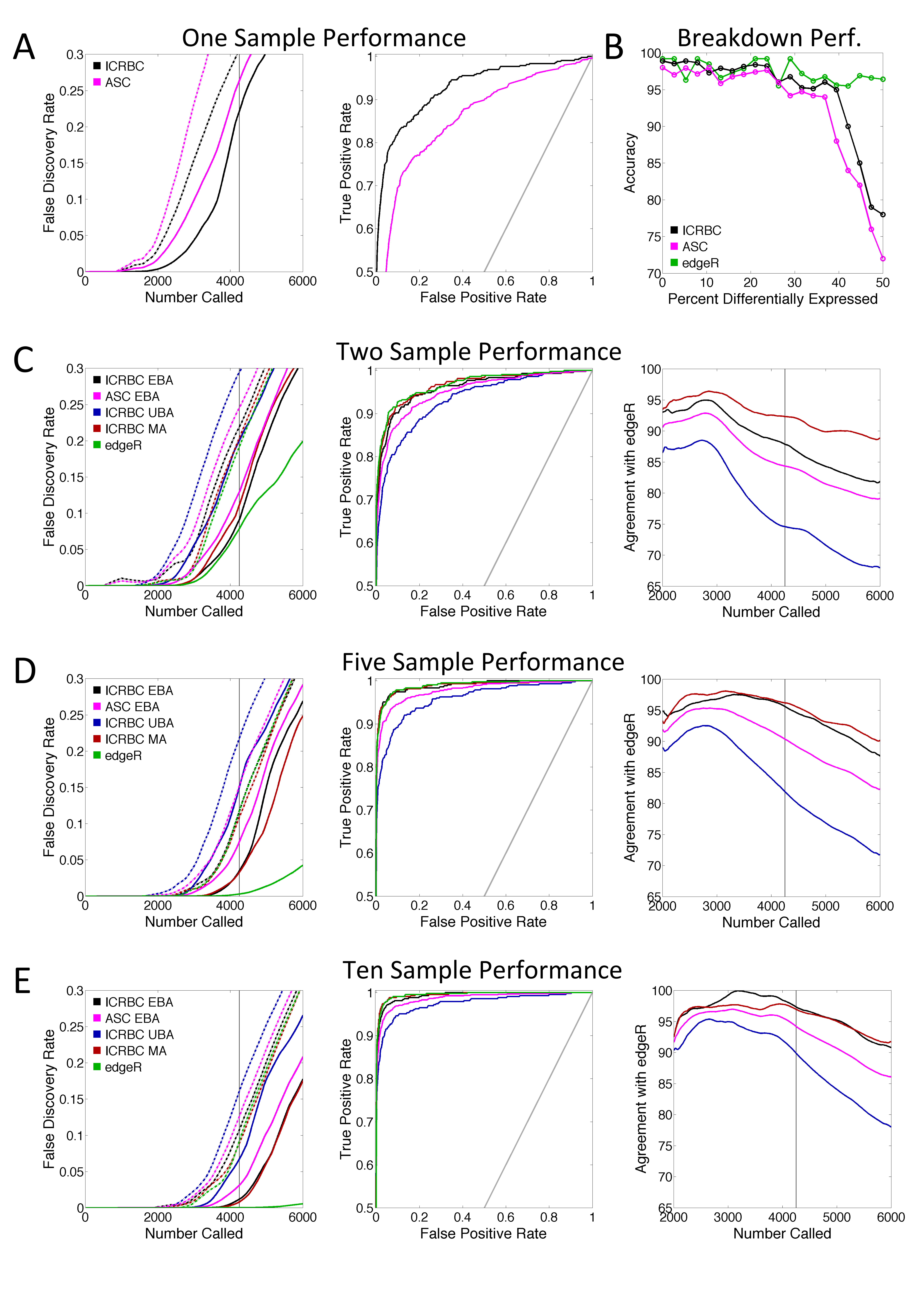}
\end{center}
\caption{
{\bf `Mathematical versus Biological Averaging' and `Breakdown' simulation study results. }  A) Performance trends of ICRBC and ASC when only a single sample is available. Left: FDR curves depicting the expected (dotted) and true (solid) FDR of each method as an increasing number of genes are called differentially expressed. Middle: ROC curves illustrating classification accuracy.  B) `Breakdown' performance of each method as depicted by their classification accuracy as a function of the percent of genes differentially expressed. Parts C-E) continue the `Mathematical versus Biological Averaging' simulation study results and show 2, 5 and 10 sample performance trends for each method, respectively. As in A) the left and middle panels illustrate FDR and ROC curves, respectively. The right panel illustrates agreement with edgeR as an increasing number of genes are called differentially expressed. Here method $x$'s agreement  with method $y$ is defined to be $|DE_n(x) \cap DE_n(y)|\div|DE_n(y)|$, where $DE_n(a)$ denotes the set of differentially expressed genes for method $a$ when $n$ genes are called.
}
\label{fig4}
\end{figure}

\textbf{Results} -- Figure \ref{fig4} illustrates the average performance of ICRBC, edgeR, and ASC on 20 independent simulation replicates. DESeq's performance is omitted here for clarity; in all cases DESeq's curves closely mimicked those of ICRBC-MA (red). ICRBC was run on not only the evenly (EBA, black) and unevenly (UBA, blue) pooled samples, but also gene-by-gene arithmetic averages of the count tables produced from the mathematically averaged experiment (MA, red). Curves for ASC illustrate it's performance on the evenly pooled biologically averaged (EBA) samples.

Figure \ref{fig4}A shows the performance of ASC and ICRBC on a single sample.  Power is relatively low, and around 2000 genes can be called differentially expressed at an expected 5\% false discovery rate. For ICRBC, the true number of false discoveries remains at or below five for 1844 calls. In other words, with a single sample ICRBC detects 1844/4260 = 43.3\% of differentially expressed genes while incurring five false discoveries. ASC detects 1054/4260 = 24.7\% of differentially expressed genes while incurring five false discoveries.  

Figure \ref{fig4}C depicts two sample performance. When ICRBC is run on the uneven biological average, it performs slightly better than a single sample, detecting 1887/4260 = 44.3\%  of differentially expressed genes while incurring five false discoveries. However, when tissue from both individuals is more evenly shared in the pool, classification improves markedly such that 55.7\% of differentially expressed genes are correctly detected at the same FDR. In fact, when running ICRBC on evenly pooled data, classification accuracy is comparable to edgeR, and interestingly to ICRBC's performance on arithmetically averaged count tables. However, edgeR consistently provides a lower true FDR. In contrast to ICRBC, ASC falls intermediate in FDR estimates and classification accuracy to ICRBC's uneven pool performance and even pool performance, suggesting it is less efficient.

The right panel in \ref{fig4}C also shows that the ordering of loci by significance using ICRBC and ASC on well pooled samples is 93\%-96\% concordant with edgeR for the top 2900 features. When considering the top 4160 loci, only ICRBC's significance ordering when run on mathematically averaged read count vectors exceeds 90\%, though ICRBC and ASC's significance ordering when run on evenly pooled samples is not far behind. ICRBC's significance ordering when run on unevenly pooled samples shows considerable disagreement with edgeR.

Five sample and ten sample performance trends (Figure \ref{fig4}C and \ref{fig4}D) are similar. ICRBC's performance on unevenly pooled samples lags behind, followed by ASC's performance. On five evenly pooled samples ICRBC classifies as well as edgeR, but is slightly outperformed when ten samples are available. However, ICRBC's classification accuracy on averaged read count vectors is not statistically different from that of edgeR's. Notice now that edgeR provides a substantially lower true FDR than all other ICRBC and ASC runs. When ten samples are available edgeR detects 99.4\% of differentially expressed features with fewer than five false discoveries. ICRBC and ASC on evenly pooled samples detect 93.2\% and 75.6\% of differentially expressed features, respectively, with five false discoveries. Because true FDR curves are not observed in practice, it is worth noting that all methods consistently overestimate the FDR.

\subsubsection*{Breakdown}

Both ASC and ICRBC assume that \emph{most} loci are not affected by alternate conditions of interest, but what is the practical limit of ``most?'' In this experiment, we examine at what percentage of truly differentially expressed genes the assumptions of high interpopulation similarity breakdown.

\textbf{Data generation} -- Read count vectors for this experiment was generated exactly as described for the `Biological versus Mathematical Averaging' experiment; however, instead of inducing a fixed number of genes for `abnormal' individuals, we induced between 0\% and 50\% at 20 evenly spaced intervals. 

\textbf{Results} -- Figure \ref{fig4}B illustrates the breakdown performance of edgeR, ICRBC, and ASC. ICRBC and ASC were run on a single `normal' and `abnormal' sample each obtained by biologically averaging from 10 evenly pooled individuals. edgeR was run on 10 `normal' and 10 `abnormal' read count vectors. 

Both ASC and ICRBC give accurate results when 0\% to 20\% of all loci are induced, though ICRBC slightly outperforms ASC in classification accuracy. When more than 25\% of all features are induced ASC and ICRBC give less reliable results than those obtained using edgeR in a mathematically averaged design.

\subsection*{Real Data}

In order to better understand how ICRBC's perfomance may compare to the other baseline methods in practice, we ran ICRBC, ASC, edgeR, and DESeq on the Cumbie, Katz, and Bottomly datasets. Note that these studies followed a mathematically averaged design, and no biologically averaged samples are available. Therefore, we ran ICRBC on either individual sample pairings between control and treatment groups (``Singles'') or on gene-by-gene arithmetically averaged count tables of the control and treatment count tables (``MA''). ASC was run exclusively on ``MA'' samples.

 ``Single'' sample pairings can be interpreted as biologically averaged samples with only a single individual enrolled, and thus provide a \emph{lower bound} for the expected performance of ICRBC. By contrast, ``MA'' runs  provide an upper bound on the expected performance of ASC and ICRBC. This is because gene-by-gene arithmetic averages of count tables are slightly better than perfectly pooled biologically averaged samples as they are also averaging over technical variation. For the Cumbie and Katz datasets all possible pairings between individual samples were considered since there were only three and two biological replicates of each condition, respectively. Because the Bottomly dataset contained ten or more biological replicates per condition, we randomly selected two from each for the ``Singles'' comparisons.

\begin{figure}[!ht]
\begin{center}
\includegraphics[width=6in]{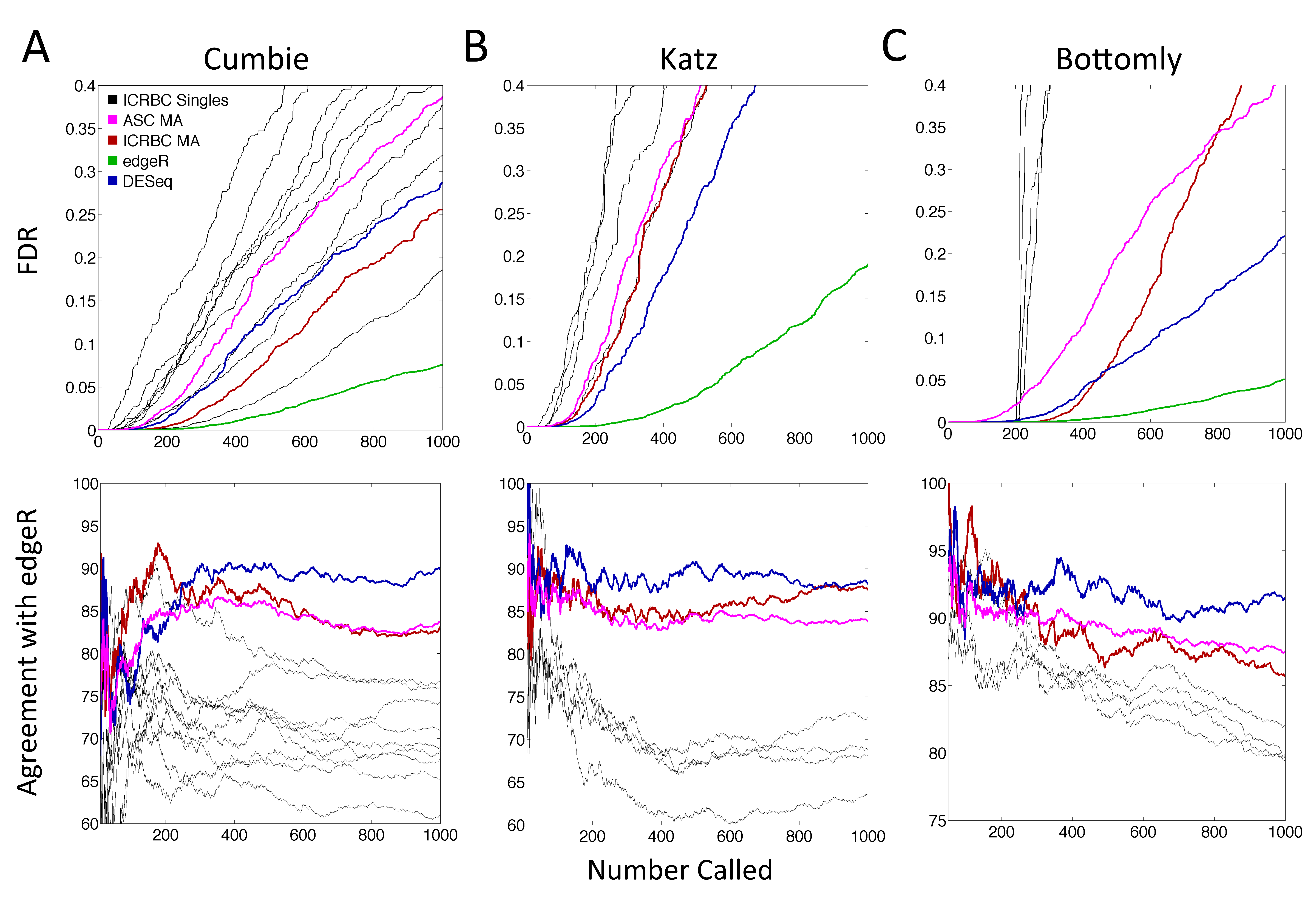}
\end{center}
\caption{
{\bf Performance trends for the A) Cumbie, B) Katz, and C) Bottomly datasets.} FDR curves (top) illustrate the behavior of the expected FDR of each method as a function of the number of genes called differentially expressed. Agreement plots (bottom) depict each methods agreement with edgeR as defined for Figure \ref{fig4}.
}
\label{fig5}
\end{figure}

The top panels in Figure \ref{fig5} illustrate how the expected FDR for each method behaves as more features are called differently expressed. The expected FDR curve for ICRBC-MA is bounded by the FDR curves of edgeR and DESeq, which tend to be liberal and conservative classifiers, respectively \cite{Soneson2013}. By contrast the FDR curve for ASC-MA majorizes the DESeq curve, which may suggest it is being too conservative. In the Katz and Bottomly datasets, both ICRBC-MA and ASC-MA FDR curves tend to lie above DESeq's. Taken together, this suggests that in an evenly pooled biologically averaged experiment, ICRBC and ASC may tend to conservatively estimate the set of differentially expressed genes.

ICRBC-MA shows considerable concordance with edgeR (Figure \ref{fig5}, bottom). Both agree on 88\%-94\% of the top 200 features in each dataset, and agreement stays above 83\% for the top 1000 features. While edgeR significance ordering may not be considered as ground truth, our simulation results and edgeR's usage of all sequenced samples, suggest it will be, on average, more accurate than ICRBC or ASC. Therefore, ICRBC-MA's agreement with edgeR implies that many of ICRBC's detected features are likely truly altered in expression.

However, these results must be taken in context of the variability and relatively higher discordance of the ``Singles'' curves seen in the Cumbie and Katz datasets. Especially in the Cumbie dataset, the ``Singles'' FDR and agreement curves illustrate the decreased reliability of highly uneven biological averaging. Even in the Katz dataset where the ``Singles'' FDR curves lie in closer proximity to each other, their agreement with edgeR decreases substantially as more features are called differentially expressed. 

Interestingly, for the Bottomly dataset, agreement with edgeR for the top 100 loci is at least 87\% for all ``Singles.'' Together with the rapid increase in the ``Singles'' FDR curves around 200 features called, this result suggests that the top 200-300 alternatively expressed loci in the Bottomly dataset may have been extractable from just a few samples. 

\section*{Discussion}

If cost isn't a limiting factor in experimental design, choosing a mathematically averaged design is ideal. However, when biological variance is large and sequencing and library preparation costs exceed tissue isolation costs, biological averaging may be more cost effective and still afford statistically tractable data. This result is intuitively reasonable because biological averaging controls exclusively for biological variation, and fewer samples are ultimately sequenced. 

When, in addition, fewer than 20\% of the regulome is expected to be differentially expressed one may use ICRBC to analyze biologically averaged experiments. Because regulatory networks are sparse, alteration of any non-essential pathway component will, in many cases, involve less than 10\% of the organism's regulome \cite{Yeung2002}. Therefore, ICRBC's 20\% breakdown threshold is reasonable for many studies, including the Cumbie, Katz, and Bottomly experiments examined here. 

If 10 or more individuals are evenly pooled in a biologically averaged design, our simulation results suggest that ICRBC may accurately detect 75\% of differentially expressed feature at an FDR as low as 0.001. Obtaining 10-20 individuals is often experimentally reasonable and within the optimal range where biological averaging is expected to be statistically more cost efficient than mathematical averaging, even if pooling is slightly uneven. Additionally, ICRBC's concordance with edgeR remains above 95\% for the top 4160 features in simulation and above 88\% for the top 400 features in the real datasets. In sum, these observations imply that ICRBC can, with substantially fewer sequenced samples, uncover the top few hundred differentially expressed loci in evenly pooled biologically averaged experiments that enroll a reasonably high number of individuals.

Nevertheless, ICRBC's and ASC's subpar performance on unevenly pooled individuals in simulation and their ``Singles'' performance in real data, clearly illustrate the potential pitfalls of highly uneven pooling -- low accuracy, and high variability. For biological averaging to be effective, care must be taken when pooling tissue samples prior to library preparation.

Finally, ICRBC's seemingly superior classification performance over ASC in simulation and greater agreement with edgeR in real data is noteworthy. Both ICRBC and ASC qualitatively define differential expression as unexpectedly large differences in log-expression given average expression; however, ASC makes heavily parametric assumptions about the distribution  expression values \emph{across} loci. In order to condition the null variance of log-expression difference on average expression, ASC assumes that $\log_{10}$RPM expression across loci follows a shifted exponential distribution, a monotonically decreasing distribution. Visual inspection of most histograms illustrating expression distributions across loci usually reveals considerable bimodality (see Figure 1 in SI), and thus argues against using a shifted exponential distribution.

ICRBC, by contrast, estimates the heteroscedastic variance function of log-expression difference using a nonparametric LOESS smoothing procedure. Given most genomes and exomes have tens to hundreds of thousands of loci, the nonparametric fitting is well supported and likely better captures regulome-wide expression distributions, which may be heterogeneous across datasets.

\section*{Conclusion}

When obtaining RNA-seq replicates is difficult, either due to cost or scale of study, biological averaging can be a useful alternative to mathematical averaging. When high interpopulation similarity is suspected and pooling evenness can be guaranteed, a biologically averaged experiment may be performed and analyzed with our ICRBC method. Biological averaging and ICRBC are not meant to replace mathematical averaging and associated analysis tools; however, biological averaging and subsequent analysis with ICRBC can provide statistically reasonable results with less than half of the data previously required. This enables biologists to enroll more individuals or, at the very least, perform reliable exploratory analyses for a fraction of the cost.

\section*{Acknowledgments}
We would like to thank Jan Prins for his helpful comments in developing the statistical theory and ICRBC method.

\bibliography{reflib}

\section*{Figures}

\newpage
\section*{Tables}
\begin{table}[!ht]
\begin{center}
  \begin{tabular}{| c | c | c | c | c |}
\hline
        		& $\hat{\alpha}$ 	& $q$ 	& $s$ 	& median depth	\\ \hline
	Cumbie 	&  0.083			& 17 	& 222 	& 5,082,539 	\\
	Katz 	&  0.047			& 15		& 193 	& 1,956,745		\\
	Bottomly 	&  0.038			& 20 	& 203 	& 3,102,573 	\\ \hline

  \end{tabular}
\caption{Estimated ERCE parameters for the three real datasets. Here $\hat{\alpha}$, $q$, and $s$ denotes the estimated dispersion parameter, tissue isolation cost, and sequencing cost, respectively. The median depth is the sum of the median (exemplar) expression profile of counts.}
\label{table1}
\end{center}
\end{table}

\end{document}


\maketitle

\tableofcontents

\section{Relative Cost Efficiency}

Let $X_{gi}$ be the true expression level of gene $g$ in individual $i$, and suppose that across individuals each $X_{g,i}$ are IID with some population level density $f_X(x)$. Recall that $O_{ij}$ denotes the observed number of reads that map to loci $i$ in individual $j$. Let the point estimate for the expression level for gene $g$ in a mathematically averaged experiment be $d_m(g,O)$, and the point estimate for a biologically averaged experiment be $d_b(g,O)$. Here $O$ is the $p$-loci by $n$-individuals count matrix. \\

For a mathematically averaged experiment, the variance of $d_m(g,O)$ is given by,
\begin{align*}
\textrm{V}[d_m(g, O)] & = \textrm{V}\left(\frac{1}{n}\sum_{i=1}^n O_{gi} \right)  \\
& = \frac{1}{n^2} \sum_{i=1}^n \textrm{V}(O_{gi}) \\
& = \frac{1}{n^2} \sum_{i=1}^n (\sigma_T^2(g) + \sigma_B^2(g))\\
& = \frac{1}{n} (\sigma_T^2(g) + \sigma_B^2(g))\\
\end{align*}
Likewise, for a biologically averaged experiment,
\begin{align*}
\textrm{V}[d_b(g, O)] = \textrm{V}(O_{g1})  & = \sigma_T^2(g)  + \textrm{V}\left(\sum_{i=1}^{n} a_i X_{gi} \right) \\
& = \sigma_T^2(g)  + \sum_{i=1}^{n} a_i^2 \textrm{V}(X_{gi}) \\
& = \sigma_T^2(g)  + \sigma_B^2(g) \sum_{i=1}^{n} a_i^2\\
& = \sigma_T^2(g)  + \sigma_B^2(g) \epsilon
\end{align*}
where $\sum a_i = 1$ and $\epsilon =\sum a_i^2$. \\

The relative efficiency of these two procedures is computed as the ratio of their variances. Thus,
\begin{align*}
\textrm{eff}_{rel}(d_b, d_m) & = \frac{\textrm{V}[d_b(g, O)]}{\textrm{V}[d_m(g, O)]} \\
& = \frac{ \sigma_T^2(g) + \sigma_B^2(g) \epsilon }{ \frac{1}{n} (\sigma_T^2(g) + \sigma_B^2(g)) } > 1
\end{align*}

To calculate the relative cost efficiency, we multiply the relative efficiency by the procedural cost ratio:
\begin{align*}
\textrm{eff}_{rc}(n | d_b, d_m) & = \frac{\textrm{cost}(d_b)}{\textrm{cost}(d_m)} \times \frac{\textrm{V}[d_b(g, O)]}{\textrm{V}[d_m(g, O)]} \\
& =  \frac{nq + s}{n(q + s)} \times \frac{ \sigma_T^2(g) + \sigma_B^2(g) e }{ \frac{1}{n} (\sigma_T^2(g) + \sigma_B^2(g)) }\\
& =  \frac{nq + s}{q+ s} \times \frac{ \sigma_T^2(g) + \sigma_B^2(g) e }{ \sigma_T^2(g) + \sigma_B^2(g) }
\end{align*}
The function $\textrm{cost}(d_i)$ gives the cost of the point estimate provided by measurment procedure $i$. 

\section{Expected Relative Cost Efficiency}

After assuming that read counts follow a $\textrm{Negative-Binomial}$ distribution, we found in the main text that the expected relative cost efficiency was given by, 

\begin{align*}
\textrm{E}[\textrm{eff}_{rc}(n, \mu | d_b, d_m) ]& =  \left(\frac{nq + s}{q + s} \right) 
\int_{\mu \in \mathbb{R}_+} \frac{ \mu + \alpha \mu^2 \epsilon }{ \mu + \alpha \mu^2 } f(\mu) \hspace{1mm}\textrm{d}\mu.
\end{align*}

\begin{figure*}[t]
\centerline{\includegraphics[scale=0.34, trim=0cm 6cm 0cm 5cm]{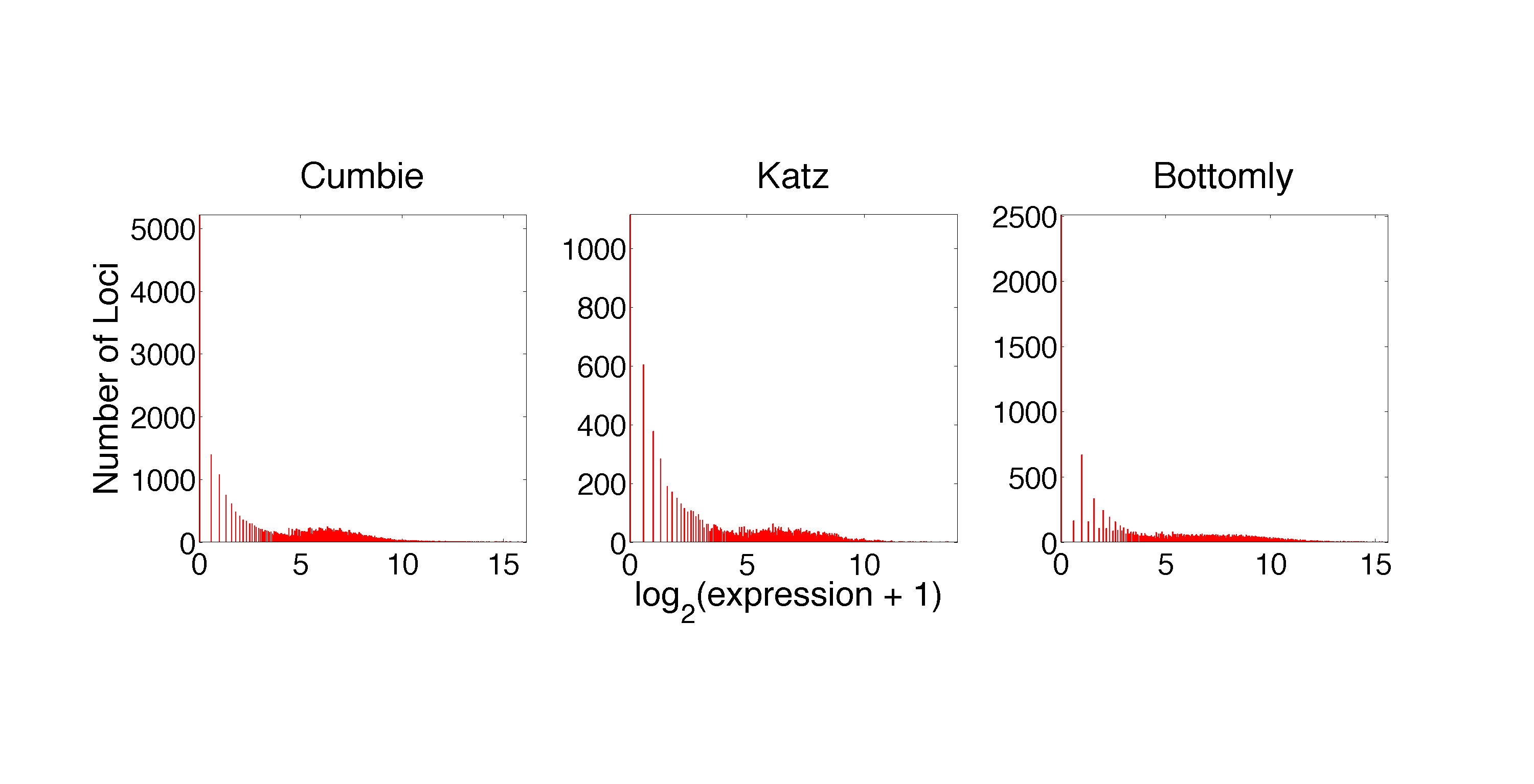}}
\caption{ Log expression distributions of the Cumbie, Katz, and Bottomly datasets. These histograms were made using 400 bins. }
\label{ed}
\end{figure*}

To our knowledge we are not aware of a probability density function that describes the distribution of expression levels for loci in a genome. We have noticed that often many, if not most, genes have low levels of expression (fewer than 30 reads), while the remaining genes are expressed at more detectable levels. Figure \ref{ed} illustrates the logarithm of the median expression distribution of the control (wildtype) samples in each of the real datasets we considered in this paper.  While  an exponential or zipf distribution could be used to describe the untransformed expression distribution, its unclear to us why, mechanistically speaking, gene expression levels across loci should necessarily follow an exponential decay pattern. \\

So without assigning a parametric family to $f(\cdot)$, the following algorithm may be used to empirically estimate $\textrm{E}[\textrm{eff}_{rc}(n, \mu | d_b, d_m) ]$ given a RNA-Seq count matrix. 

\subsection{Empirical Estimation of the Expected Relative Cost Efficiency}

\begin{enumerate}
\item Given a $p$-loci by $n$-individuals count matrix $O$,  remove loci (rows) of $O$ that do not have any reads in any samples (0's all the way across the table).
\item Generate 100 bootstrap replicate tables of $O$ (subsample $p$ rows of $O$ with replacement 100 times to produce 100 ``new'' tables).  
\item Run edgeR on each boostrap replicate table, and let the estimate of $\alpha$ be the average of edgeR's \texttt{common.dispersion} output statistic from each run. Call the bootstrapped estimate of $\alpha$, $\hat{\alpha}$.\footnote{Though bootstrapping will give a stable estimate of $\alpha$, in our experience the $\alpha$ estimated from each replicate rarely varies by more than 5\%. For computational ease, boostrapping can probably be skipped. Instead the original table, $O$, may be used as an inpute to edgeR.}
\item Choose the number of individuals for which the expected relative cost efficiency should be evaluated. Call this number $k$. 
\item Select a pooling evenness:
\begin{itemize}
\item If \emph{perfectly even}, set $\epsilon$ to $1/n$.
\item If \emph{perfectly uneven}, set $\epsilon$ to $1$.
\item If \emph{uniformly uneven}, generate a random vector, $a$, of uniform random numbers of length $k$ and divide $a$ by its sum. Set $\epsilon$ to $\sum_{i=1}^{k}a_i^2$. In the main text we generated multiple uniformly uneven pooling weights so that we could visually assess the variance associated with uniformly uneven pooling.
\end{itemize}
\item Set a tissue isolation cost $q$, and library preparation plus sequencing cost, $s$. We estimated these costs from the methods reported in the Cumbie, Katz, and Bottomly papers with respect to today's (\today) prices. A table of our estimates are given in the main text, and we provide a discussion of how we obtained these estimates in the next section.
\item Calculate the median expression profile to be the vector whos $i^{th}$ entry denotes the median expression level of the $i^{th}$ loci across all samples within a particular population (e.g. all wildtype samples).
\item Organize the loci in this median expression profile by expression level (counts) into $B$ ordered bins (just like the histogram in Figure \ref{ed}, but untransformed). We arbitrarily set $B$ to 400. However, we get statistically indistinguishable results for $100 \leq B \leq 1000$ (data not shown).
\item Let $u(i)$ be the average expression level of all loci that fall into the $i^{th}$ smallest bin, and let $|B(i)|$ denote the number of loci that fall into the $i^{th}$ smallest bin. Nonparametrically estimate the expected relative cost efficiency using the following formula,
\begin{align*}
\textrm{E}[\textrm{eff}_{rc}(k, \mu | d_b, d_m) ]& \approx  \left(\frac{kq + s}{q + s} \right)
\sum_{i = 1}^{B} \frac{ u(i) + \hat{\alpha} u(i)^2 \epsilon }{ u(i) + \hat{\alpha} u(i)^2 } \cdot \frac{|B(i)|}{p}.
\end{align*}
\end{enumerate}

\begin{figure*}[t]
\centerline{\includegraphics[scale=0.28, trim=0cm 2cm 0cm 0cm]{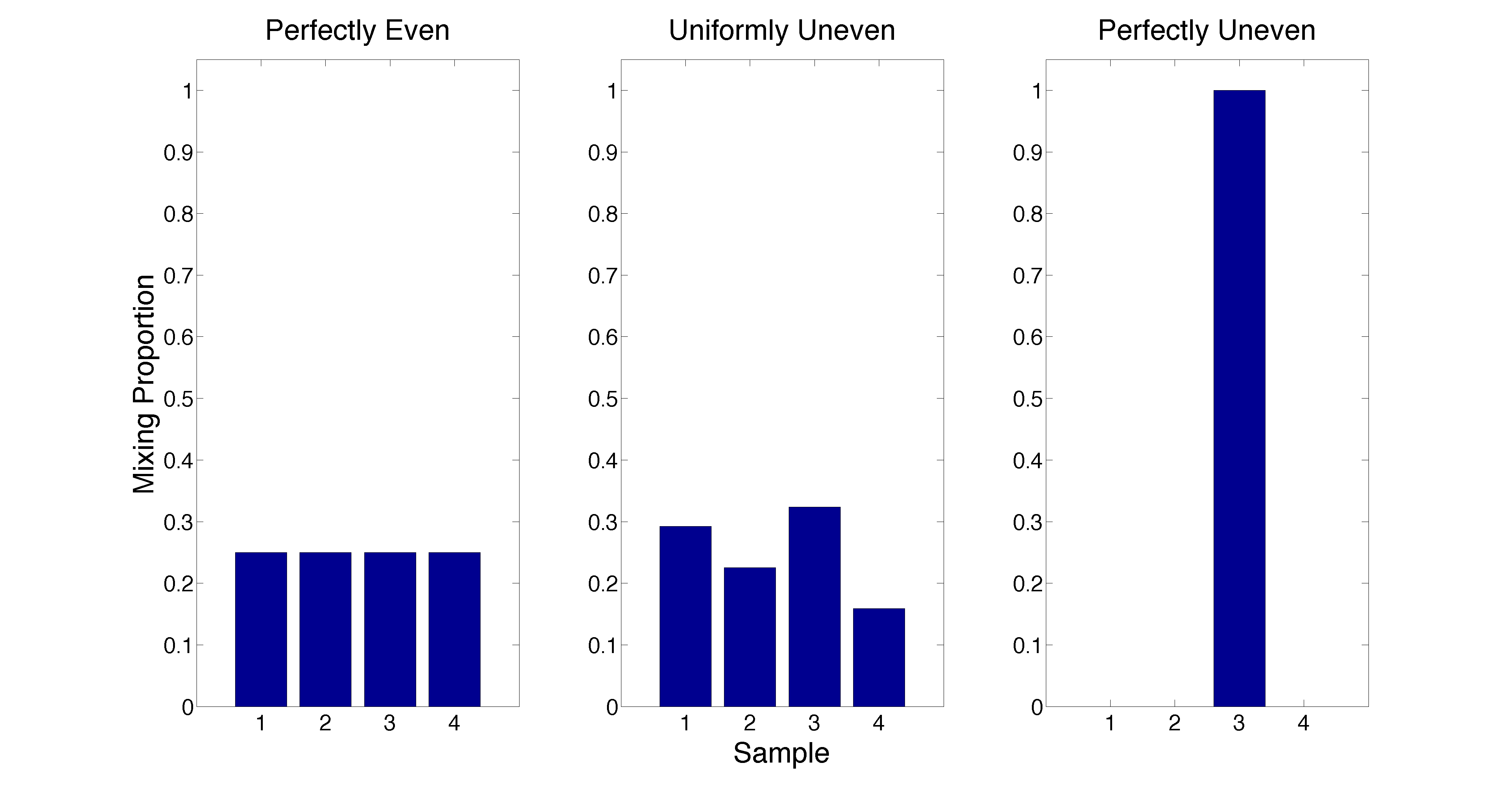}}
\caption{ Pooling types considered in ERCE estimation for the Cumbie, Katz, and Bottomly datasets. Perfectly even and perfectly uneven pools represent the best and worst case, respectively. Uniformly uneven pools are more likely to represent reality. }
\label{pooltypes}
\end{figure*}

Pooling evenness, $\epsilon$, was assumed to be either perfectly even, perfectly uneven, or uniformly uneven (Figure \ref{pooltypes}). Uniformly uneven pooling weights for $n$ individuals are obtained by generating an $n$-long Uniform[0,1] random vector and normalizing it by its sum.

\subsection{Cost estimation for ERCE comparisons on the real datasets}

Tissue isolation cost, $q$, involved the cost of TRIzol reagent (Invitrogen, Carlsbad CA), RNAlater (Qiagen, Valencia CA), and system specific costs. We did not include the cost of RNA quantification equipment (e.g. Agilent Bioanalyzer, Santa Clara, CA) in this assessment. The libarary preparation and sequencing cost, $s$, was estimated using prices from the High Throughput Sequencing Facility (\texttt{https://sites.google.com/site/htsfunc/}) at the University of North Carolina at Chapel Hill. As of this writing, the price per library is \$175.61 and a TCGA standard (~165 million reads per lane on an Illumina HiSeq 2500)  RNA-Seq run is \$1502.60. We assumed that the \$175.61 was a fixed cost per individual enrolled; however, we assumed that each individual's library could be perfectly multiplexed in a single lane with enough external samples such that only the desired depth for that individual sample was obtained. Thus, to calculate the cost of obtaining another sequenced sample at the median depth, $d$, obtained from each study, we used
\[
s = 175.61 + \underbrace{d\times1502.6/(165 \times 10^6)}_{\textrm{cost of a single sample at depth $d$}}
\]

For example, Bottomly \emph{et al} require scalpels and petri dishes (system specific costs ~\$12), and TRIzol reagent and RNA later (~\$8 for just the volume they report they use). Together, their tissue isolation cost was therefore $q = 20$ dollars. The median read depth in the Bottomly study was 3,102,573. Thus, their library preparation and sequencing cost, $s$, amounts to $175.61 + 1956745\times 1502.6 / (165 \times 10^6 ) \approx 203$. 

\section{Approximate normality of M given A}

Let $O_1$ and $O_2$ denote the observed counts of a given gene in two different samples. Assume that $O_1$ and $O_2$ are independent. Let $p_1$ and $p_2$ denote the true probability a read will map to this gene in each respective sample, and let $n_1$ and $n_2$ denote the total number of reads in each respective sample. The Random Sampling model for RNA seq asserts that the sequencing process samples completely at random a limited amount of all of the total transcript available in a tissue. This can be seen (and has been shown ref) as a binomial experiment, such that $O_i \sim \textrm{Binomial}(n_i, p_i)$. Given that $n_i$ is typically large and $p_i$ is typically small (since there are many genes), it is often assumed (to good approximation) that, $O_i \sim \textrm{Poisson}(\mu_i)$, where $\mu_i = n_ip_i$. \\

The Random Sampling model has been use satisfactorily to describe technical variation; however, the Poisson assumption cannot adequately account for biological variation. The Negative Binomial 2 (NB2) model is a generalization of the Poisson model, and is capable of handling overdispersion. As a Poisson-Gamma mixture model, the NB2 distribution has expected value and variance given by $\textrm{E}[X] = \mu = np$ and $\textrm{V}[X] = \mu + \alpha\mu^2 = np + \alpha(np)^2$, where $n$ and $p$ are the parameters of the underlying binomial process, and $\alpha$ is the dispersion parameter measuring the extent of biological variation. \\

We assume that $O_i \sim \textrm{NB2}(\mu_i, \alpha)$. Let $K_i = \log(O_i)$, and define $M = K_1 - K_2$ and $A = (K_1 + K_2)/2$. To show that the conditional distribution of M given A is normally distributed, we must first show that each $K_i$ are asymptotically normally distributed. Our proof closely follows (DEGseq SI ref).

As $n_1$ and $n_2$ grow large, as is typical for a sequencing experiment, the asymptotic distribution of $O_i/n_i$ is given by,
\[
\sqrt{n_i}\left( \frac{O_i}{n_i} - p_i \right) \overset{D}{\rightarrow} \mathcal{N}(0, p + \alpha n p^2)
\]
We can equivalently write this as,
\[
\sqrt{n_i}\left( O_i - \mu_i \right) \overset{D}{\rightarrow} \mathcal{N}(0,  \mu_i + \alpha \mu_i^2)
\]
Using Delta Method, we can obtain the asymptotic distribution (as $n_i \rightarrow \infty$) of $K_i$ as follows,
\begin{align*}
\sqrt{n_i}( K_i - \log(\mu_i) ) & = \sqrt{n_i}(\log(O_i) - \log(\mu_i)) \\
& \overset{D}{\rightarrow} \mathcal{N}\left(0, (\mu_i + \alpha\mu_i^2) \left[ \frac{\partial}{\partial\mu_i} \log(\mu_i)\right]^2 \right) \\
& = \mathcal{N}\left(0, \frac{1 + \alpha\mu_i }{\mu_i} \right)
\end{align*}
Therefore we have, 
\[
K_i \overset{D}{\rightarrow} \mathcal{N}\left(\log(\mu_i), \frac{1 + \alpha\mu_i }{n_i\mu_i} \right)
\]
At this point we have shown that $K_i$ is approximately normally distributed for large $n_i$. For convenience let $\tau_i  = \log(\mu_i)$ and $\sigma_i^2  = (1 + \alpha\mu_i )/(n_i\mu_i)$. Given that $O_1$ and $O_2$ are independent, $K_1$ and $K_2$ will be independent. Therefore, we have 
\begin{align*}
M & \sim \mathcal{N}\left( \tau_1 - \tau_2, \sigma_1^2 + \sigma_2^2 \right) = \mathcal{N}\left( \tau_M, \sigma_M^2\right) \\
A & \sim \mathcal{N}\left( \frac{1}{2}(\tau_1 + \tau_2), \frac{1}{4}(\sigma_1^2 + \sigma_2^2) \right) = \mathcal{N}\left( \tau_A, \sigma^2_A \right)  \\
\end{align*}

Finally, the conditional distribution of a normal random variable given another is also a normal random variable (ref). So, the conditional distribution of $M$ given $A = a$ is,
\[
M |_{A = a} \sim \mathcal{N}\left( \tau_M + \rho\frac{\sigma_M}{\sigma_A}(a - \tau_A), \sigma_M^2(1 - \rho^2) \right)
\]
Here $\rho$ is the correlation coefficient between $M$ and $A$.

\subsection{Invariance of the mean of $M$}
The expected value is invariant with respect to $A$ under the null distribution that $p_1 = p_2 = p$. 
\begin{align*}
\textrm{E}[M] & = \tau_1 - \tau_2 \\
&= \log(\mu_1) - \log(\mu_2) \\
& = \log(n_1p_1) - \log(n_2p_2) \\
& = \log(n_1p) - \log(n_2p) \hspace{5mm} \textrm{(under nullity)}\\
& = \log(n_1) - \log(n_2)
\end{align*}
This final expression depends only on the sequencing depths of the two samples, which are both known at the time of analysis. Thus, the mean of $M$ does not depend on $A$. 

\subsection{Heteroscedasticity with respect to $A$}
To clearly see the heteroskedastic dependence of the variance of $M$ on $A$, 
\begin{align*}
\sigma_M^2 &= \sigma^2_1 + \sigma^2_2 \\
&= \frac{1 + \alpha\mu_1}{n_1 \mu_1} + \frac{1 + \alpha\mu_2}{n_2 \mu_2} \\
&= \frac{n_2\mu_2 + n_1\mu_1 +\mu_1\mu_2 \alpha(n_1 + n_2)}{n_1n_2\mu_1\mu_2} \\
&= \frac{n_2\mu_2 + n_1\mu_1 + \exp\left\{(\log(\mu_1) + \log(\mu_2)\right\} \alpha(n_1 + n_2)}{n_1n_2\exp\left\{(\log(\mu_1) + \log(\mu_2)\right\} } \\
&= \frac{n_2\mu_2 + n_1\mu_1 + \exp\left\{\textrm{E}[K_1] + \textrm{E}[K_2]\right\} \alpha(n_1 + n_2)}{n_1n_2\exp\left\{\textrm{E}[K_1] + \textrm{E}[K_2]\right\}} \\
&= \frac{n_2\mu_2 + n_1\mu_1 + \exp\left\{\textrm{E}[K_1+K_2]\right\} \alpha(n_1 + n_2)}{n_1n_2\exp\left\{\textrm{E}[K_1+K_2]\right\}} \\
&= \frac{n_2\mu_2 + n_1\mu_1 + \exp\left\{2\textrm{E}[A]\right\} \alpha(n_1 + n_2)}{n_1n_2\exp\left\{2\textrm{E}[A]\right\}}
\end{align*} 

In an MA plot we can visually see that the variance of $M$ decreases in increasing $A$. Does the above expression agree with this intuition? To answer this, it will be sufficient to show that the derivative of $\sigma_M^2$ with respect to $A$ is negative,

\begin{align*}
\frac{\partial\sigma_M^2}{\partial A} &= -\frac{2\exp(-2\textrm{E}[A])(\mu_1n_1 + \mu_2n_2)}{n_1n_2}Af(A)
\end{align*} 
where $f(A)$ is the exact density of $A$. This expression is in fact negative, which agrees with our intuition.

\section{Pooling weights for the simulation study}
\begin{figure*}[!t]
\centerline{\includegraphics[scale=0.35, trim=0cm 2cm 0cm 0cm]{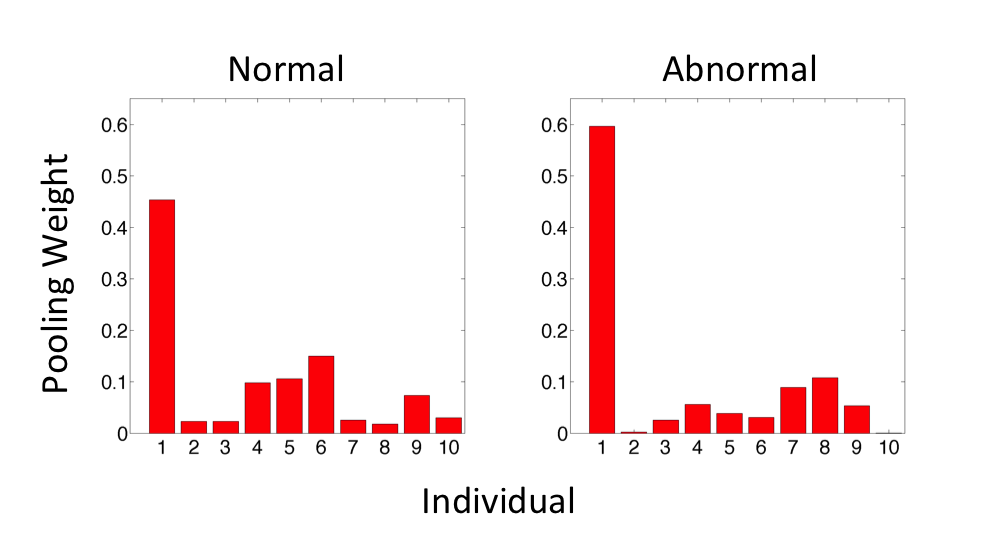}}
\caption{ Pooling weights for simulated `normal' and `abnormal' individuals in the `Biological versus Mathematical Averaging' and `Breakdown' simulation studies. }
\label{pooltypes}
\end{figure*}

In the `Biological versus Mathematical Averaging' and `Breakdown' simulation studies, we considered even or uneven transcript vector pools. For uneven pooling, we wanted to consider a more uneven pool than is likely to occur in practice. \\

Figure \ref{pooltypes} illustrates the pooling weights for individuals that were unevenly pooled. When considering uneven pooling weights for an $n$ sample performance comparison in the `Biological versus Mathematical Averaging' experiment, the first $n$ pooling weights in each treatment group (`Normal' and `Abnormal'), were taken and divided by their sum. Notice that the first individual in `Normal' and `Abnormal' samples always represented more than 40\% of the pool.

%
%